# SOLVING THE INVERSE PROBLEM FOR ARTIFICIAL IRREGULARITY DIAGNOSTICS FROM EXTRAORDINARY WAVE ANOMALOUS ATTENUATION DATA

N.A. Zabotin, A.G. Bronin, G.A. Zhbankov

*Rostov State University, Rostov-on-Don, Russia*

**Abstract.** The paper presents formulation and solution for the problem of determination of spatial spectrum of artificial ionospheric irregularities from vertical HF radio sounding data on anomalous attenuation. The suggested method uses the dependence of extraordinary wave attenuation caused by multiple scattering from the irregularities on frequency to obtain the spectrum parameters. The method is applied to the data obtained in experiments with Sura heating facility carried out on September 6-9, 1999.

## Introduction

The main purpose of the ionosphere radio sounding is to determine the structure of the medium from the characteristics of the reflected signal. The formulation of corresponding inverse problem is well-known for determination of the ionosphere regular properties (like the altitude dependence of electron density $N\ (h)$ or collision frequency $\nu_e(h)$) from the vertical or oblique sounding data. The inverse problem in this case is reduced to solving of Fredholm or Volterra type integral equations.

Diagnostics of spatial spectrum of random ionospheric irregularities from parameters of the sounding signal is connected with solving of the inverse problem also. One of the signal characteristics that can be used for this purpose is non-collisional attenuation caused by multiple scattering. According to the theory of this effect, under conditions typical for non-disturbed mid-latitude ionosphere the anomalous attenuation magnitude may reach 10 dB and more [Zabotin et al., 1998]. In the present paper we consider the frequency dependence of anomalous attenuation as the input parameter for the inverse problem of the irregularity spectrum determination.

The essential peculiarity of formulation of the inverse problem in the case of the diagnostics of random ionospheric irregularities is the complexity of direct problem. The solution of the direct problem depends upon various parameters of the irregularity spectrum and propagation geometry. And this dependence is rather complex. This complexity does not allow to formulate the inverse problem in analytic form, for example, as a system of differential or integral equations. In this case to solve the inverse problem means to find such parameters of the model of radio waves propagation in random medium which give the best (in the sense of certain norm in functional space) fit of experimental data. Let us suggest that $L^{\exp}(x)$ is the observed dependence of some characteristic of the signal, reflected from the ionosphere, on parameter $x$ and $L(x,\mathbf{p})$ is the same dependence calculated from the model of the medium, defined by the set of parameters $\mathbf{p}=\{p_1,p_2,...\}$. The solution of inverse problem is to find such set of model parameters $\mathbf{p}$ which gives the minimal deviation of $L(x,\mathbf{p})$ from $L^{\exp}(x)$:

$$\min\left\|L^{\exp}(x)-L(x,\mathbf{p})\right\|.$$

The matter of direct problem in this case is determination of spatial-angular distribution of radiation reflected from the ionospheric layer for given regular parameters of the layer (electron density profile, etc.) and given parameters of spatial spectrum of irregularities. For the case of vertical sounding the dependence of signal attenuation $L$ on frequency is used as input experimental data.

This approach to solving of inverse problem for radio sounding of irregular ionospheric layer was suggested and tested on experimental data in [Bronin et al, 1999].

In the present paper we describe application of this method to diagnostics of artificial irregularities for the case of extraordinary sounding wave.

In Section 1 we describe the method of direct problem solving. The formulation of inverse problem is discussed in Section 2. Experimental data and obtained results are discussed in Section 3.

## 1. Formulation and solution of direct problem

The ray path in plane-stratified medium may be unambiguously defined by invariant ray variables – polar and azimuth angles of arrival $\theta, \varphi$ at the point $\vec{\rho}$ on the certain base plane which is parallel to the layer. It is natural to choose the surface of the Earth as the base plane for the case of ionospheric propagation. The spatial and angular distribution $P_0(\theta,\varphi,\vec{\rho})$ of radiation energy reflected from the ionosphere is described by the radiation balance equation (RBE) [Zabotin *et al.*, 1998]:

$$\frac{d}{dz}P(z,\vec{\rho},\theta,\varphi) = \int Q(z;\theta,\varphi;\theta',\varphi')\cdot \cdot\left\{P(z,\vec{\rho}-\vec{\Phi}(z;\theta',\varphi';\theta,\varphi),\theta',\varphi') - P(z,\vec{\rho},\theta,\varphi)\right\}d\theta' d\varphi' \,, \tag{1}$$

where $z$ is vertical coordinate, $Q(z;\theta,\varphi;\theta',\varphi') = \sigma(\theta,\varphi;\theta',\varphi')C^{-1}(z,\theta,\varphi)\sin\theta'\left|\frac{d\Omega'_k}{d\Omega'}\right|$, $C(z;\theta,\varphi)$ is cosine of the inclination angle of the ray path with invariant angles $\theta$ и $\varphi$; $\left|\frac{d\Omega'_k}{d\Omega'}\right|$ is the Jacobean of transition from polar and azimuth angles of wave vector to invariant angles, $\sigma(z,\theta,\varphi,\theta',\varphi')$ is the scattering cross-section. Vector function $\vec{\Phi}(z;\theta,\varphi;\theta',\varphi')$ describes the displacement of the point of arrival of the scattered ray with angle coordinates $\theta',\varphi'$ from the point of arrival of the incident ray with angle coordinates $\theta$, $\varphi$.

In the approximation of small-angle scattering in the invariant angle variables (which is not identical to common small-angle scattering approximation) the solution of REB equation is

$$P(z,\vec{\rho},\omega) = P_0\left[\vec{\rho} - \vec{D}(z,0;\omega),\omega\right], \tag{2}$$

where $P_0(\theta,\varphi,\vec{\rho})$ is the spatial and angular distribution of radiation energy reflected from ionosphere in absence of irregularities,

$$\vec{D}(z,0;\omega) = \int_0^z dz' \int d\omega' Q(z;\omega,\omega')\vec{\Phi}(z';\omega,\omega')\,. \tag{3}$$

According to (2) and (3) the scattering causes deformation of the distribution of radiation energy, while the type of distribution function is not changed. If only the one ray with the invariant angles $\theta_0(\vec{\rho})$, $\varphi_0(\vec{\rho})$ arrives in every point at the Earth surface $\vec{\rho}$ (as it takes place for the point source and frequencies below critical frequency), then the function $P_0$ has the following form

$$P_0(\vec{\rho},\theta,\varphi) = \tilde{P}_0(\vec{\rho})\delta[-\cos\theta + \cos\theta_0(\vec{\rho})]\delta[\varphi - \varphi_0(\vec{\rho})], \tag{4}$$

where the quantity $\tilde{P}_0(\vec{\rho})$ is proportional to energy flux through the point $\vec{\rho}$ in absence of scattering. Substitution of (4) into (2) gives:

$$\tilde{P}(\vec{\rho}) = \tilde{P}_0\left[\vec{\rho} + \vec{D}(\theta_1,\varphi_1)\right]\cdot\left|\frac{\partial(\rho_{0x},\rho_{0y})}{\partial(\theta,\varphi)}\right|\left|\frac{\partial(\rho_{0x}-D_x,\rho_{0y}-D_y)}{\partial(\theta,\varphi)}\right|^{-1}_{\substack{\theta=\theta_1,\\ \varphi=\varphi_1}}. \tag{5}$$

where $\vec{D}(\theta,\varphi) \equiv \vec{D}(z_0,0;\omega)$. New angles of arrival $\theta_1$ and $\varphi_1$ are found from the system of equations

$$\vec{\rho} = \vec{\rho}_0(\theta_1,\varphi_1) + \vec{D}(\theta_1,\varphi_1)\ , \tag{6}$$

where $\vec{\rho}_0(\theta,\varphi)$ is the point of arrival at the base plane of the ray with invariant ray variables $\theta$ and $\varphi$ in the absence of scattering.

According to (5), observer situated at the point $\vec{\rho}$ will notice two effects caused by scattering: the change of the angles of arrival and the attenuation of intensity of received signal, which is determined as

$$L = 10\lg \frac{\tilde{P}(\vec{\rho})}{\tilde{P}_0(\vec{\rho})}. \quad (7)$$

## 2. Formulation and solution of inverse problem

As it was explained in Introduction, to solve the inverse problem means to find such set of model parameters $p_j$, which give the minimal deviation of calculated values of the anomalous attenuation $L(\omega_i, p_1, p_2, ..., p_m)$ from experimental values measured at $n$ frequencies $\omega_i$, $i = 1,2,...n$ in the quadratic norm:

$$\min_{p_j, j=1,2,...m} \sum_{i=1}^{n} \left( L_i^{\exp}(\omega_i) - L(\omega_i, p_1, p_2, ..., p_m) \right)^2 \quad (8)$$

Therefore we have a classical nonlinear least squares problem. The methods of numeric solving of nonlinear least squares problem are well developed [Dennis and Schnabel, 1983]. Realizations of these methods are available through various numeric analysis libraries such as, for example, MINPACK [More *et al.*, 1980].

Now we need to determine the model of random irregularity spectrum to make our model of propagation in the ionosphere complete. Let us suppose that the irregularities in the given band of scales are strongly stretched along the lines of geomagnetic field force and let us characterize them by the spectrum of the following form

$$F(\vec{\kappa}) \propto \delta_R^2 (1 + \kappa_\perp^2 / \kappa_{0\perp}^2)^{-\nu/2} \delta(\kappa_\parallel), \quad (9)$$

where $\kappa_\perp$ and $\kappa_\parallel$ are respectively orthogonal and parallel to lines of geomagnetic field force components of wave vector of irregularities $\vec{\kappa}$, $\kappa_{0\perp} = 2\pi / l_{0\perp}$, $l_{0\perp}$ is upper scale of spectrum, $\delta(x)$ is delta-function. The spectrum is normalized at the value of structural function $D_N(\vec{R}) = \left\langle \left[ \delta_N(\vec{r} + \vec{R}) - \delta_N(\vec{r}) \right]^2 \right\rangle \equiv \delta_R^2$ (where $\delta_N(\vec{r}) = \Delta N / N$) for the orthogonal scale $R = 1$ km. Thus the spectrum is characterized by three parameters: $\delta$, $\nu$ and $l_{0\perp}$.

The parameters of spectrum $\delta_R$, $l_{0\perp}$, $\nu$ can evidently vary from point to point inside ionosphere. For the case of vertical sounding of plane-stratified ionosphere it is naturally to consider that these parameters depend only on altitude. Such dependence may be approximated by linear $\delta_R(h) = ah + b$ or parabolic $\delta_R(h) = ah^2 + bh + c$ dependence. This approximation may be applied either to whole ionospheric layer or to separate altitude regions in dependence of the peculiarities of experimental data.

The above model of the spectrum is not quite adequate neither for artificial irregularities nor for natural. The real spectra are known do demonstrate different behavior for different ranges of wave numbers [Szuszczewicz, 1987; Frolov *et al.*, 1996]. Real spectrum of artificial irregularities, for example, is known to have a hump in the range of scale 0.5 - 1 km. In this paper we do not intend to provide ultimate diagnostic procedure, taking into account numerous peculiarities of real irregularity spectra, electron density profile, etc. However, the valuable results may be obtained using the simple model of the above spectrum described.

## 3. Application to experimental data

The experiments with the Sura heating facility were carried out on the 6th, the 7th and the 9th of September, 1999 in the evening or night hours of the local time when absorption in the ionosphere D-region is small and may be neglected. Heating was provided by synchronous work of three 250 kW transmitters. With account of antenna gain the effective transmitted power was 300 MW. The period of each heating cycle was 5 minutes and the pause between cycles was 10 minutes. In the experiment on the 6th of September, 1999 the heating was performed by the high-power HF wave of ordinary polarization at the frequency 5,752 MHz. The attenuation of the probe wave was measured at frequencies 4.069, 4.669, 5.669, 6.069, 6.269, 6.424 and 6.849 MHz. In the experiment at the 7th of September the extraordinary heating wave with the frequency 7,815 MHz was used. Attenuation measurements were made at 8.024, 7.789, 7.624, 7.224, 6.624, 5.424 MHz. In the experiment of the 9th of September the attenuation was measured at frequencies 4.469, 4.969, 5.369, 5.569, 5.769, 5.969 and 6.169 MHz. In each case the attenuation was measured for the extraordinary waves. Diagnostic waves were emitted as a pulses with duration of 100 microseconds and with linear polarization. The reflected signals were received by the antenna tuned for extraordinary polarization with 10 dB separation of ordinary wave. During the heating experiment the profile of electron density was checked with the oblique sounding with linear FM signals at the path Yoshkar Ola – N. Novgorod.

The primary data sets (records of amplitude for seven frequencies ) were averaged over the heating cycles. The averaged curves were divided into 4 different parts. The first part corresponds to the state before the beginning of the heating, when only natural irregularities are present in the ionosphere. The second part corresponds to the period of anomalous attenuation development. The third part corresponds to saturated state when both natural and artificial irregularities are present and the fourth part corresponds to relaxation of artificial irregularities after the turning off the heating facility. The fitting was applied for each of them independently. Linear fits were used to determine saturated amplitudes before and after the beginning of the heating, while exponential fits were used to estimate the development and relaxation times of the phenomena. This method was used to obtain the dependencies of the anomalous attenuation on frequency of the probe wave (Fig. 1). It should be noted, that we determine relative attenuation of wave amplitude $L_R$, i.e. the ratio of the amplitude before the beginning of the heating to the amplitude after the beginning of the heating calculated in decibels as $L_R = 10 \cdot \lg\left(A_{before}^2 \big/ A_{after}^2\right)$, where $A_{before}$ and $A_{after}$ are the amplitudes of the wave before and after the beginning of heating.

The observed attenuation of extraordinary waves (probe waves in all series of measurements as well as pump wave at the 7-th of September) is considerably large. Multiple scattering from irregularities within the range of scales 100 m – 5 km is the only known mechanism which can explain it. Thus the significance of the obtained data is that they clearly demonstrate the existence of strong anomalous attenuation of extraordinary waves in heating experiments as well as that they indirectly prove the conclusions of multiple scattering theory.

The information available from our data sets is not enough for determination of all irregularity spectrum parameters simultaneously. Let us assign to $l_{0\perp}$ and $\nu$ rather typical values: $l_{0\perp} = 10$ km and $\nu = 2.5$. Thus we can find the dependence of $\delta_R$ on the altitude. We need to determine also the model of natural irregularities which determines background anomalous attenuation. Let us suppose that natural irregularities are described by the same spectrum (9) with the same parameters $l_{0\perp} = 10$ km and $\nu = 2.5$, and their amplitude $\delta_0$ does not depend upon the altitude. In other words we suppose that heating of ionosphere results only in the change of altitude distribution, while the type of the spectrum is not changed. In this case the quantity $\delta_0$ is a free parameter of our calculations.

To determine $\delta_R$ for the experimental value of anomalous attenuation $L_R$ one needs to determine background attenuation $L_0$ for given $\delta_0$, find the total attenuation $L = L_0 + L_R$ and then determine $\delta_R$ from the calculated dependence of the attenuation on $\delta_R$ (see illustration of this method at Fig. 2). Profiles of electron density were obtained from the ionograms of vertical sounding. For each diagnostic frequency the real profile was replaced with equivalent linear profile. To simplify numeric solution of inverse problem the approximation of isotropic ionosphere was used in calculation of quantity $\tilde{P}$.

Since we need to determine only one parameter of spectrum $\delta$ one may simplify calculations and reduce the solution of inverse problem to the solution of direct problem. Real electron density profile, determined from the ionosphere sounding data, was approximated by parabolic profile with reasonable accuracy. For each frequency of probe waves this profile was replaced by equivalent linear profile, determined from condition of equality of derivatives of the electron density over an altitude at reflection point. The attenuation of probe waves was calculated using formulae (6) and (7) for a grid of values of $\delta$. Two values of $\delta$ from the grid, giving attenuation closest to experimental value were used to start dichotomy process, giving the value of $\delta$ with required accuracy. This process was re-peated for various values of background irregularities amplitude $\delta_0$. The results of calculations are presented at Fig. 3 - 5. In this approach we replace the real ionospheric profile of electron density with the equivalent linear profiles different for different sounding frequencies. Such replacement does not cause significant error in determination of altitude dependence of $\delta$ because the main contribution into anomalous attenuation comes from the region near the point of reflection for given frequency where the linear approximation gives the best fit of real profile. Other method of calculations is to use linear or parabolic approximation of altitude dependence of $\delta$ and find the coefficients of approximation by solving the least squares problem. The results of the calculations for parabolic approximation are presented at Fig. 6 - 8. At Fig. 9 - 11 anomalous attenuation determined from restored profile of $\delta_R$ is compared with experimental values.

As it was mentioned above, the model of irregularity spectrum is rather simple and does not reflect all peculiarities of real spectra of natural or artificial irregularities. In particular, one may take into account the increase of irregularity level near the point of reflection of pump wave by the simple modification of irregularity altitude profile model

$$\delta_R = az^2 + bz + c + d\exp\left\{\frac{(z - z_H)^2}{dz_H^2}\right\},$$

where $a$, $b$, $c$ and $d$ are coefficients to determine, $z_H$ is the height of reflection of pump wave, $dz_H$ is the half width corresponding to $df = 150$ kHz. The calculations for such model were performed for the data of the 7th of September. Corresponding results are presented at Fig. 12, 13. It may be noted that account for the "hump" at the height of reflection of pump wave results in better approximation of frequency dependence of anomalous attenuation. Restoration of altitude profile of $\delta$ for both parabolic and parabolic with a hump models for September, 7 data demonstrates important peculiarity of the method. In both cases calculations give a negative values of $\delta$ for altitude in several points. It is caused by using non-uniform grid of frequencies and a relatively short set of sounding frequencies used in experiment. This problem may be solved by addition of new points using the interpolation procedure. Using of tight grid of frequencies makes such deviations not possible (see Fig. 13). The other method is to divide the whole interval of altitudes into regions within which anomalous attenuation is a monotonous function as it was done in [Bronin *et al.*, 1999].

## Conclusion

The main result obtained in the paper is demonstration of possibility of using of the anomalous attenuation measurements data for the purpose of diagnostics of artificial ionospheric irregularities. The solution of the inverse problem was performed in the simplified form which is determined by the nature of experimental data and calculation difficulties in realization of more sophisticated approach to inverse problem solution. Such approach demands, in particular, measurements on a very tight and uniform grid of frequencies or interpolation of the data. In our calculations we used the data on the anomalous attenuation of an extraordinary wave, because it is not affected by the anomalous absorption what makes possible to determine pure effect of scattering. The other reason is that existing experimental set is not able to measure anomalous attenuation for waves of both polarizations simultaneously. Simultaneous measurements of anomalous attenuation of ordinary and extraordinary waves would make it possible to separate contributions from anomalous absorption and scattering and thus to estimate the shape of spectrum for different scales of irregularities.

*Acknowledgments*. The work was supported by the Russian Foundation of Basic Research under grant No. 99-02-17525.

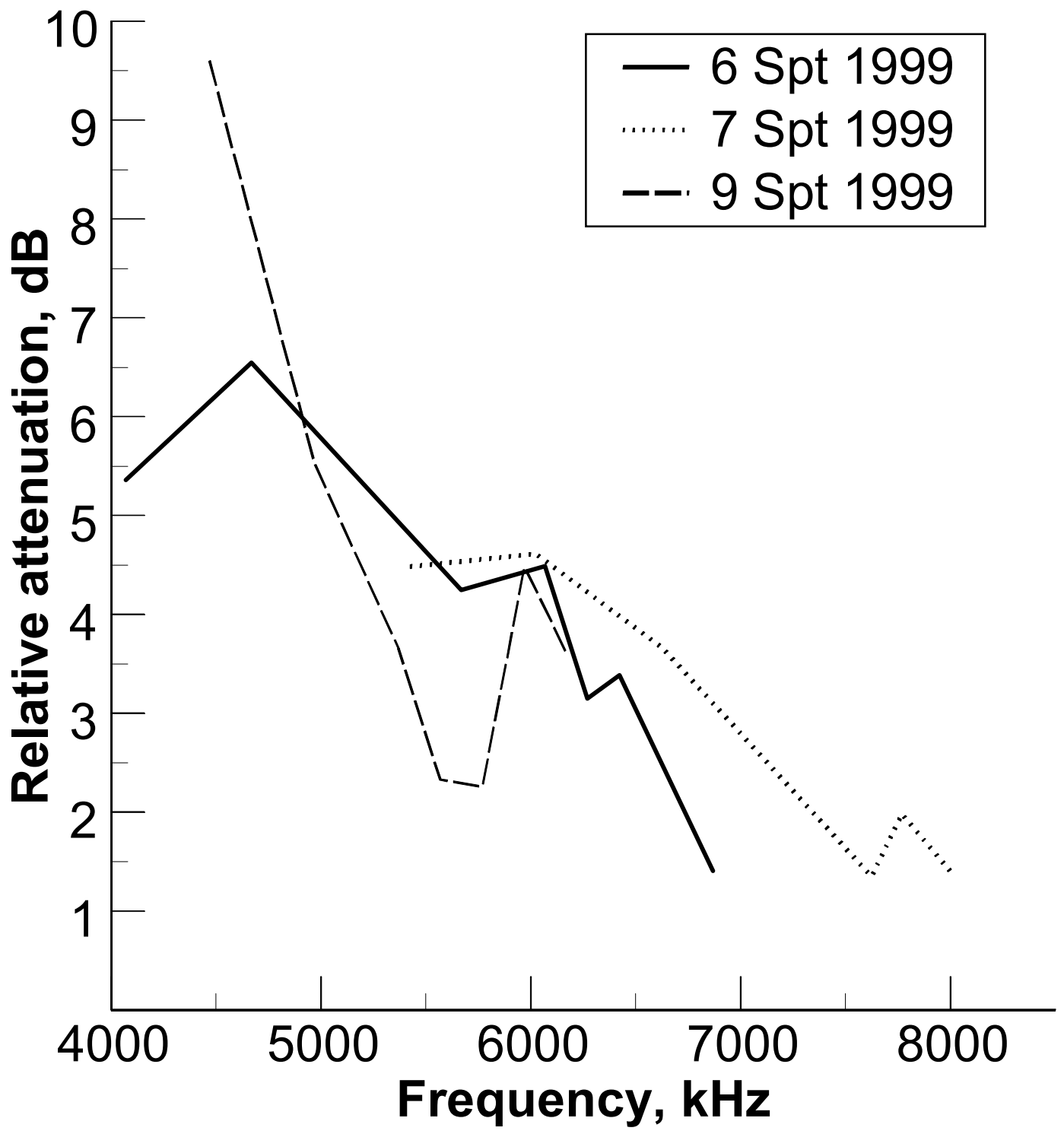

Fig. 1.
Anomalous attenuation of probe waves (excess over background value) for the 6th, the September 7 and 9, 1999.

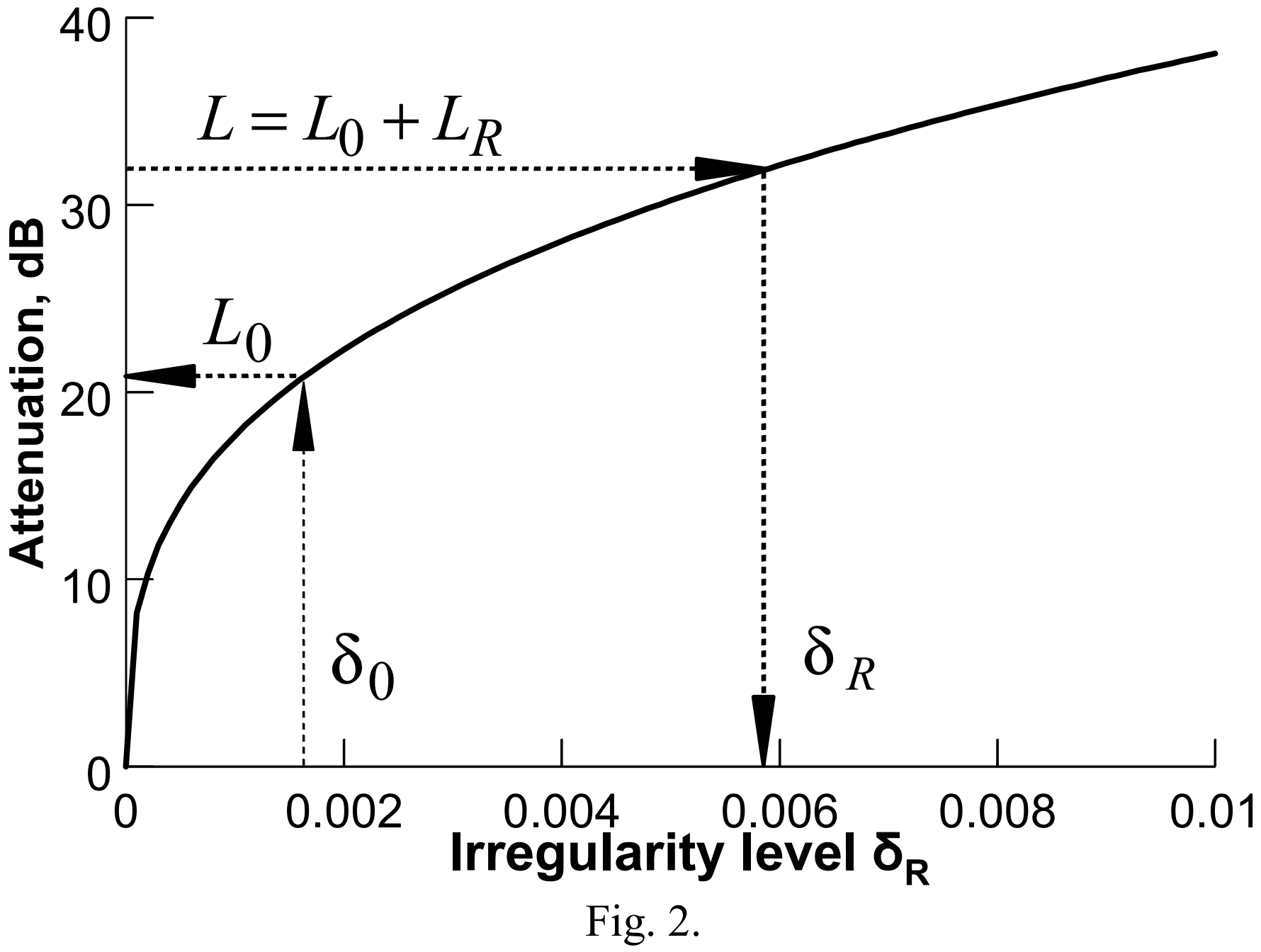

Fig. 2.
Illustration of inverse problem solution method for the case of artificial irregularities.

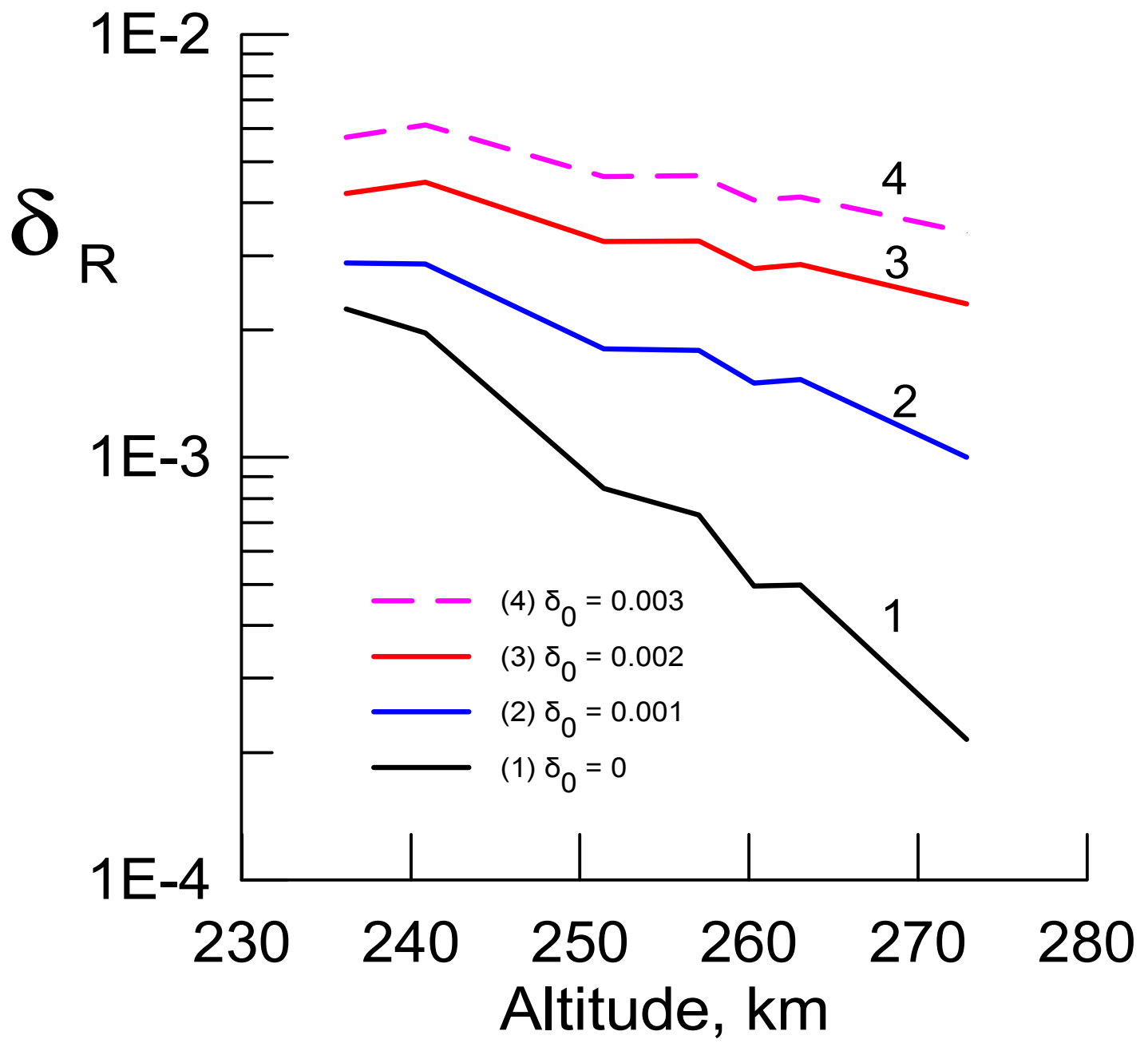

Fig. 3.
Altitude dependencies of $\delta_R$ restored by simple method for September 6, 1999 data.

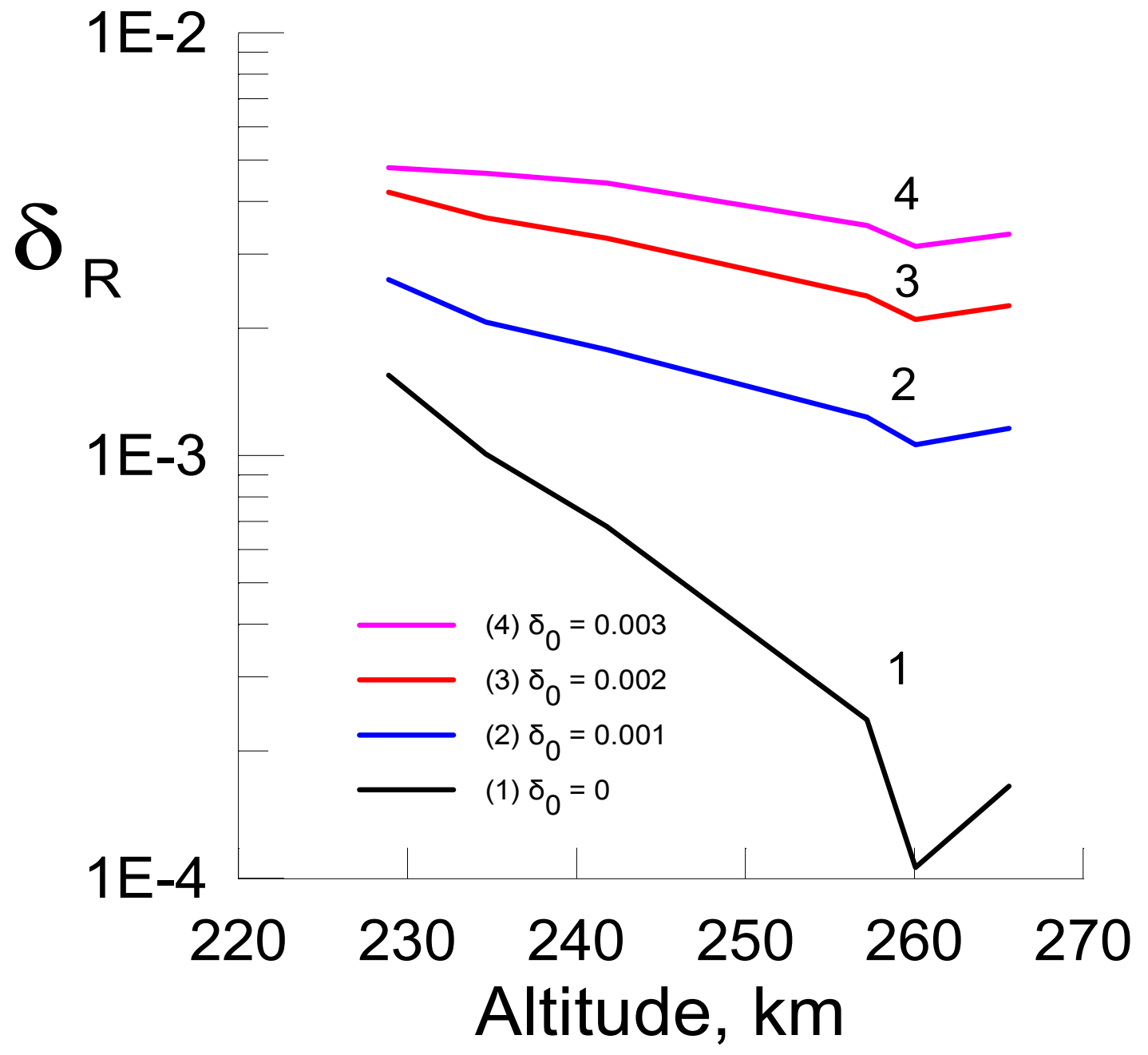

Fig. 4.
Altitude dependencies of $\delta_R$ restored by simple method for September 7, 1999 data.

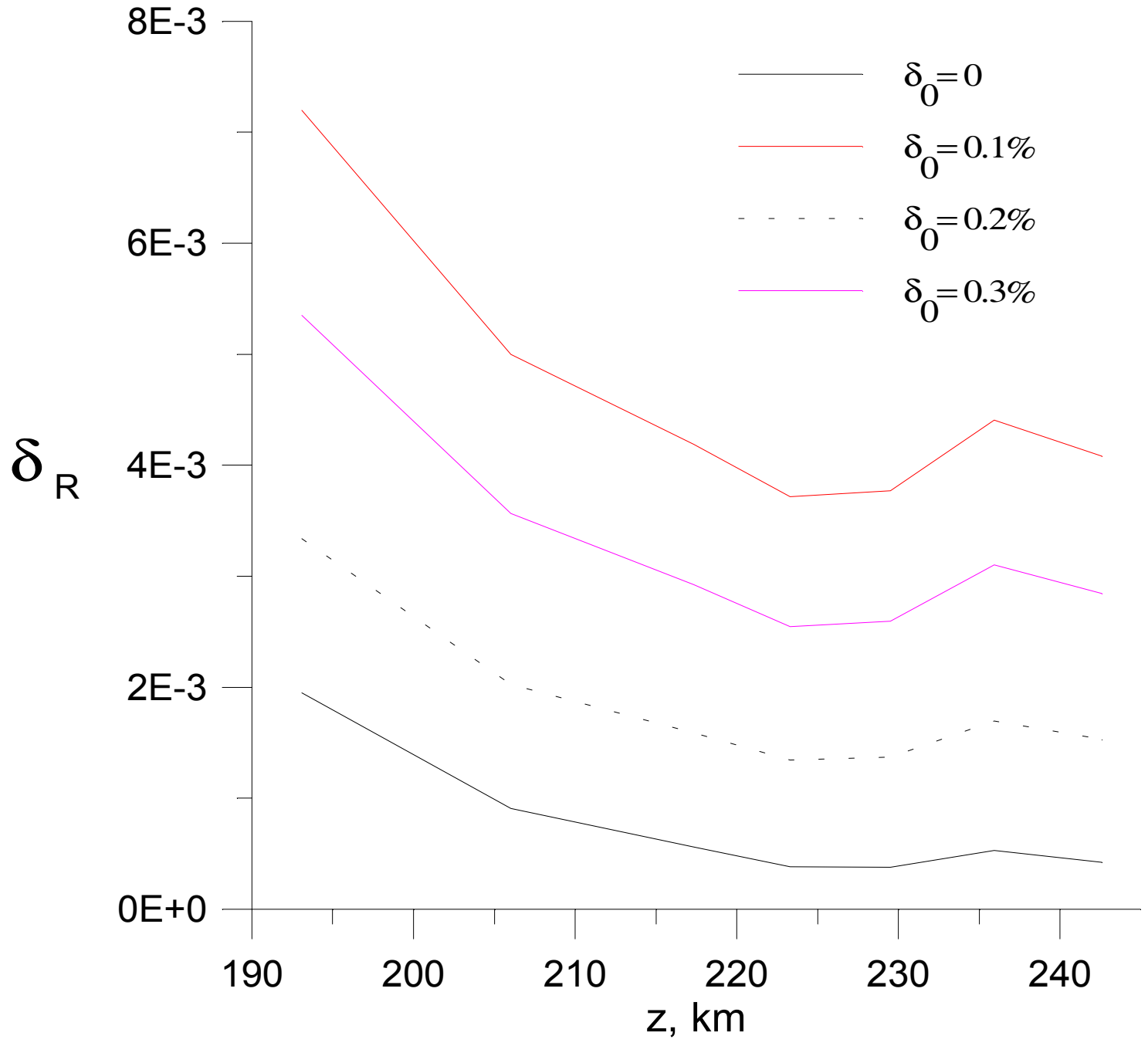

Fig. 5
Altitude dependencies of $\delta_R$ restored by simple method for September 9, 1999 data.

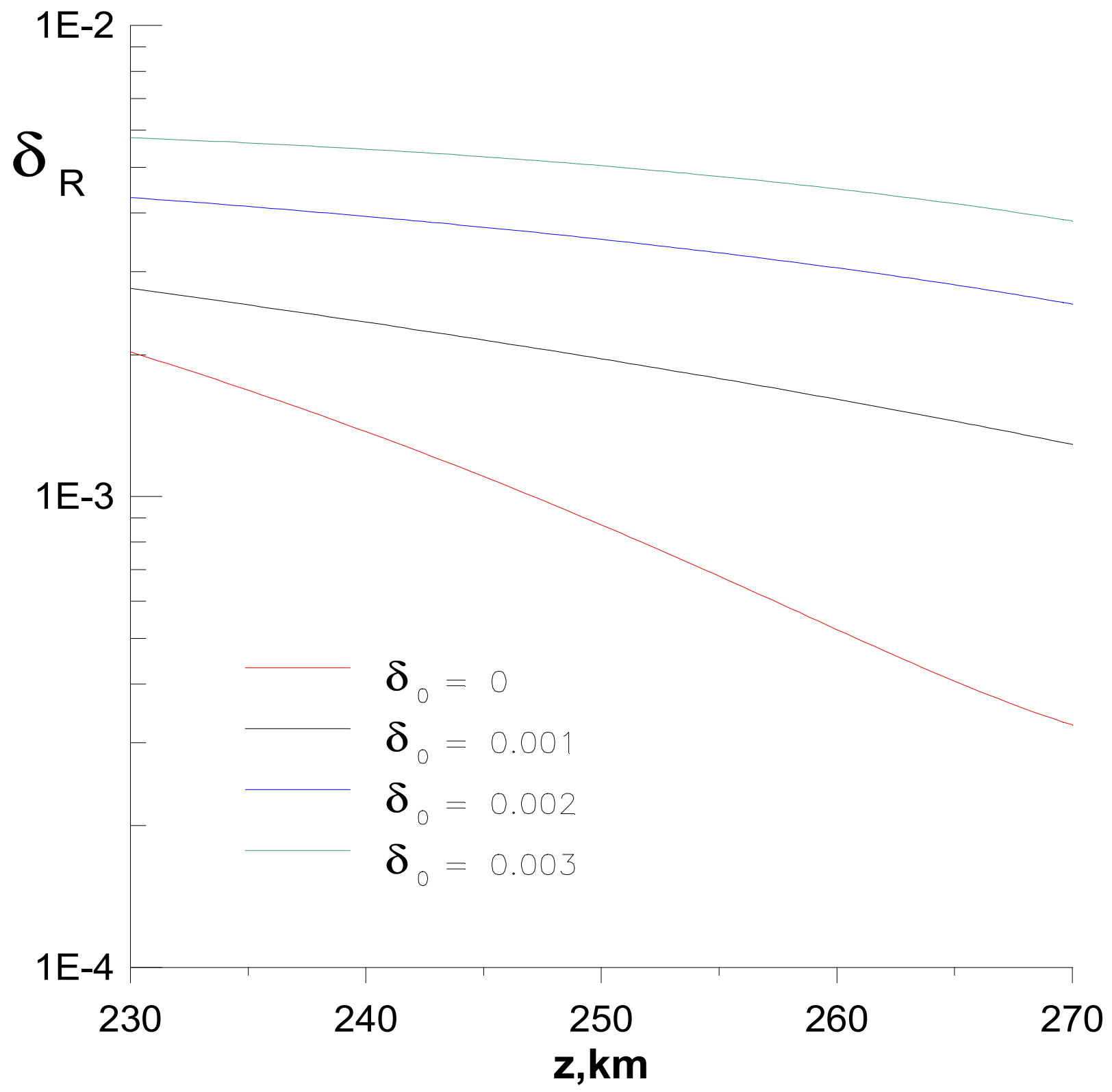

Fig. 6.
Altitude dependencies of $\delta_R$ restored for parabolic model for September 6, 1999 data.

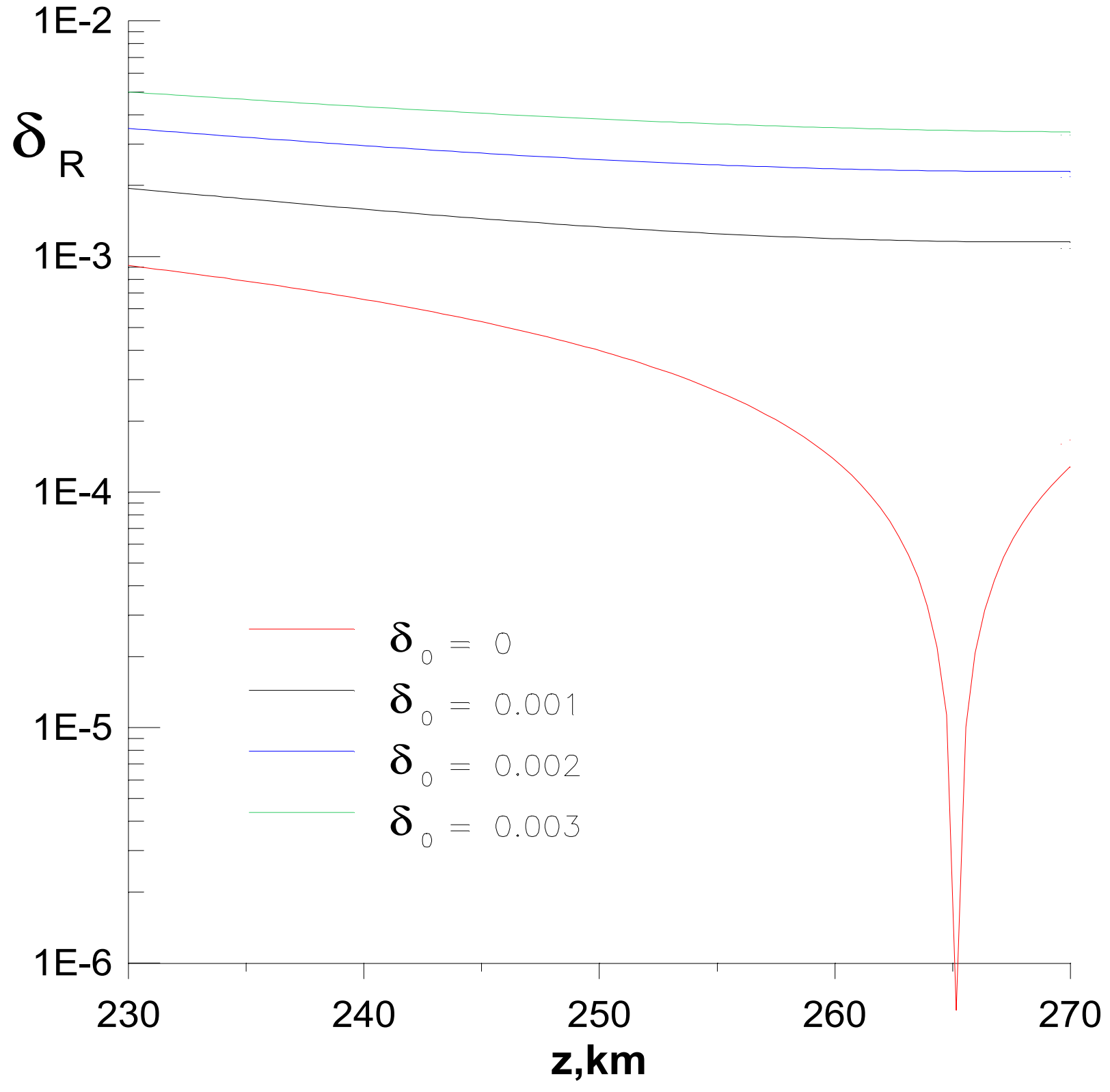

Fig. 7.

Altitude dependencies of $\delta_R$ restored for parabolic model for September 7, 1999 data.

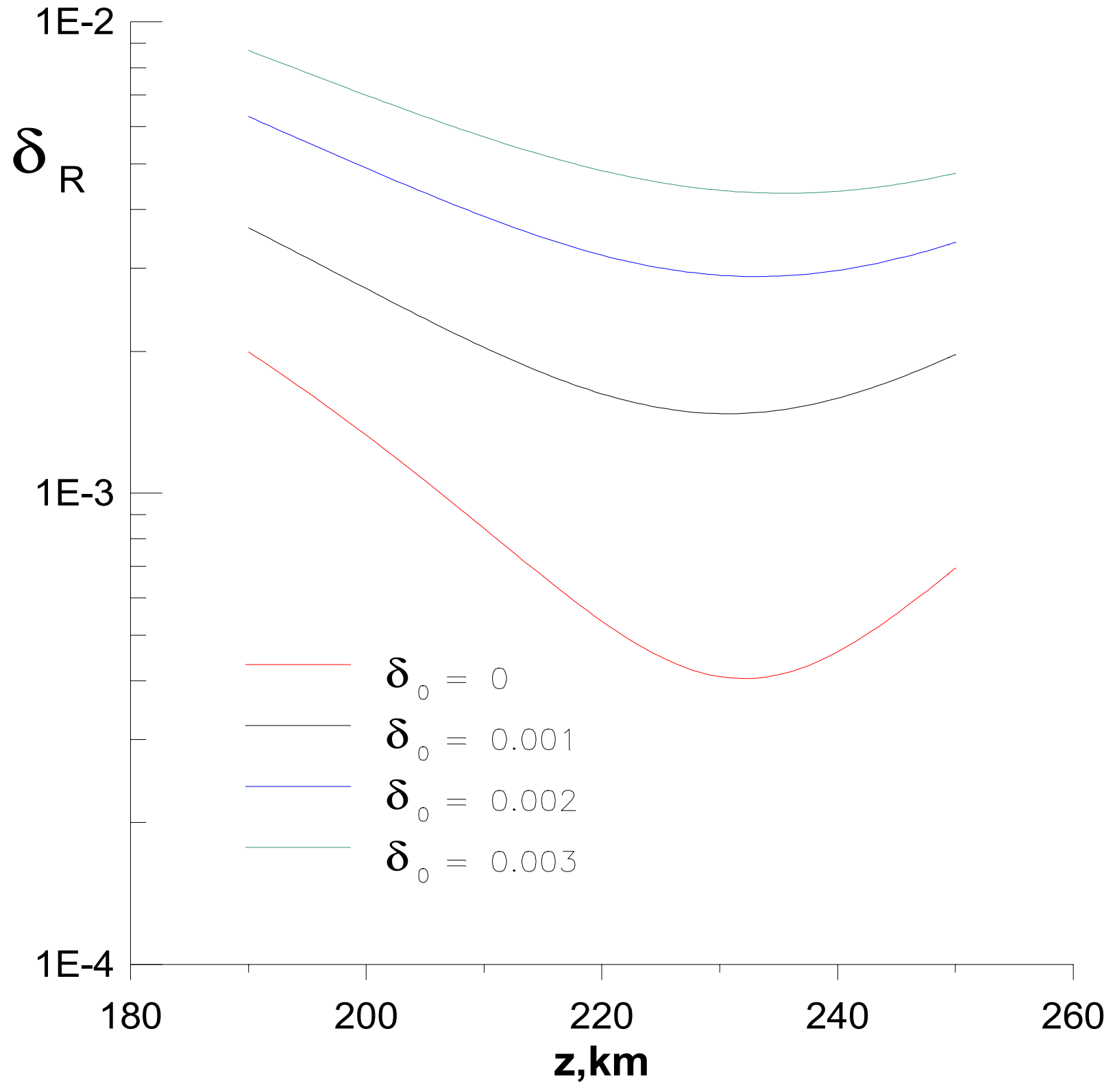

Fig. 8.

Altitude dependencies of $\delta_R$ restored for parabolic model for September 9, 1999 data.

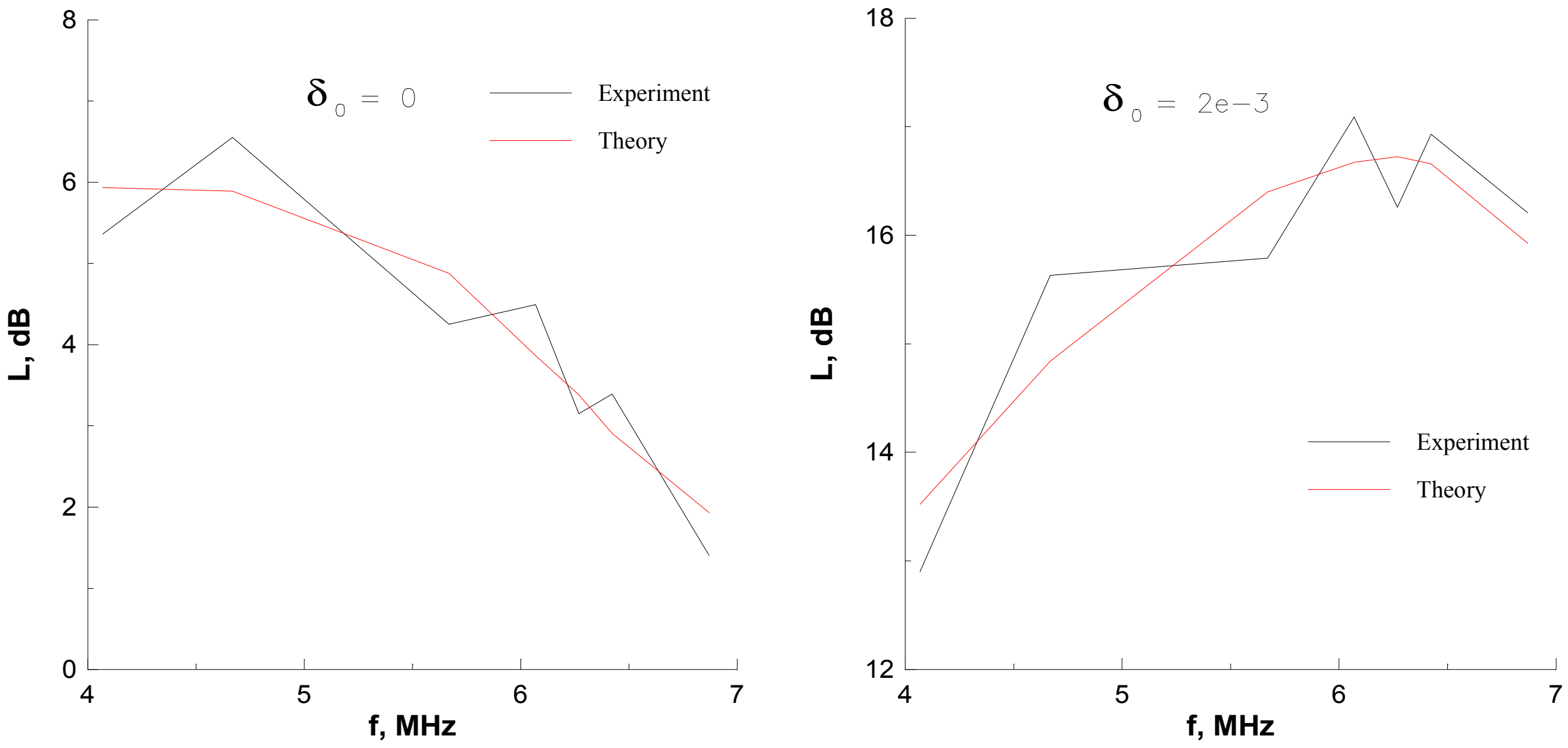

Fig. 9.
Comparison of anomalous attenuation calculated for parabolic model with experimental data for September, 6, 1999.

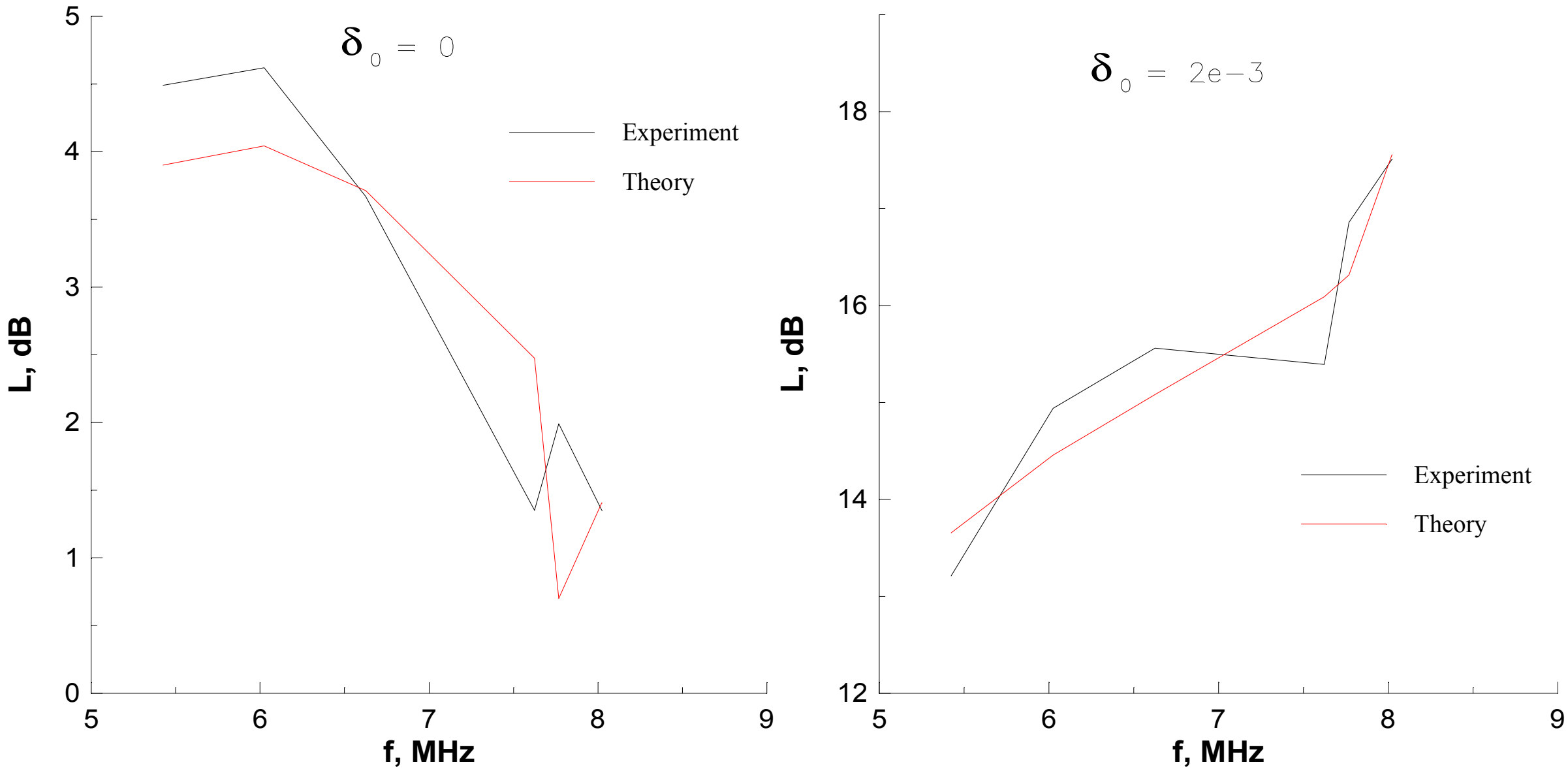

Fig. 10.
Comparison of anomalous attenuation calculated for parabolic model with experimental data for September, 7, 1999.

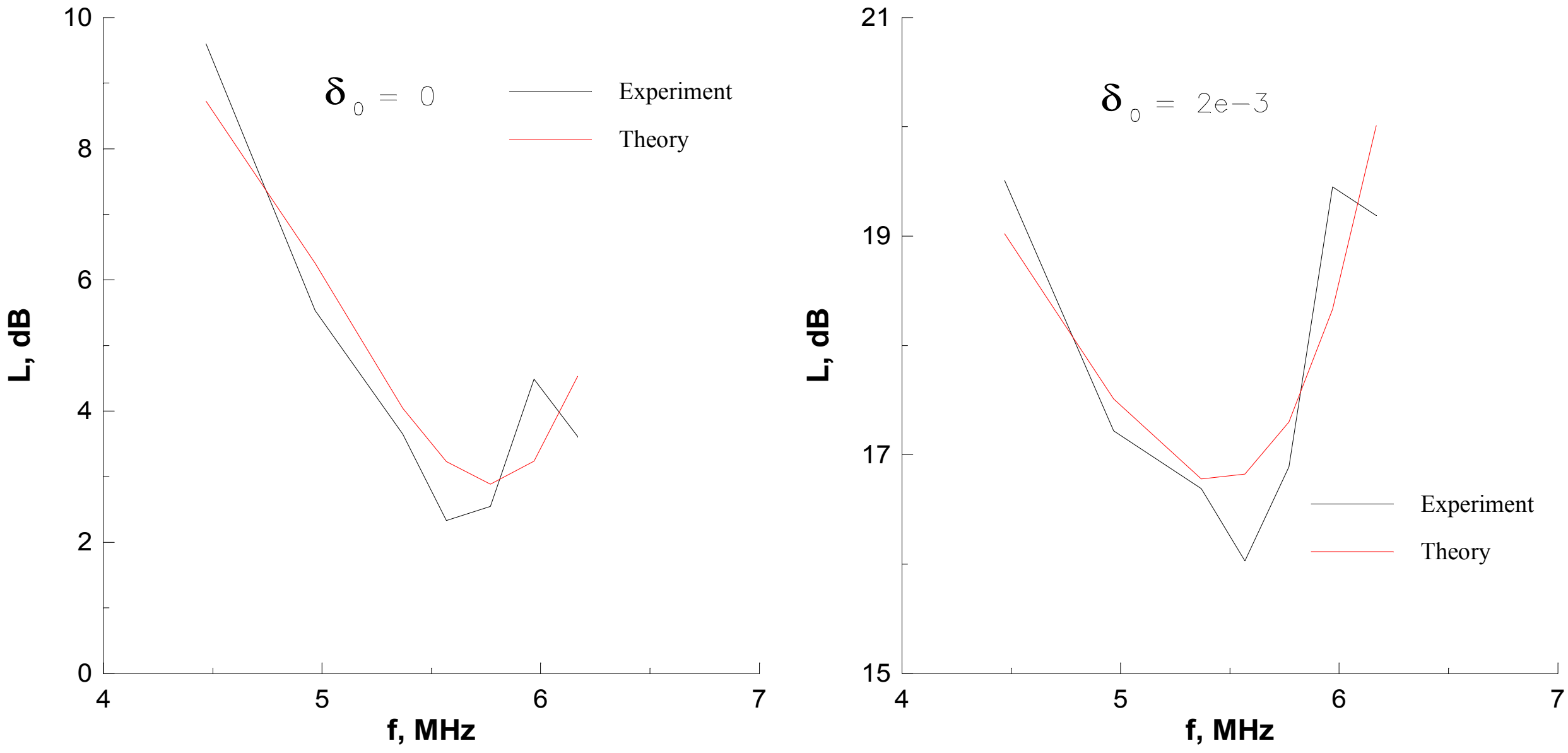

Fig. 11.
Comparison of anomalous attenuation calculated for parabolic model with experimental data for September, 9, 1999.

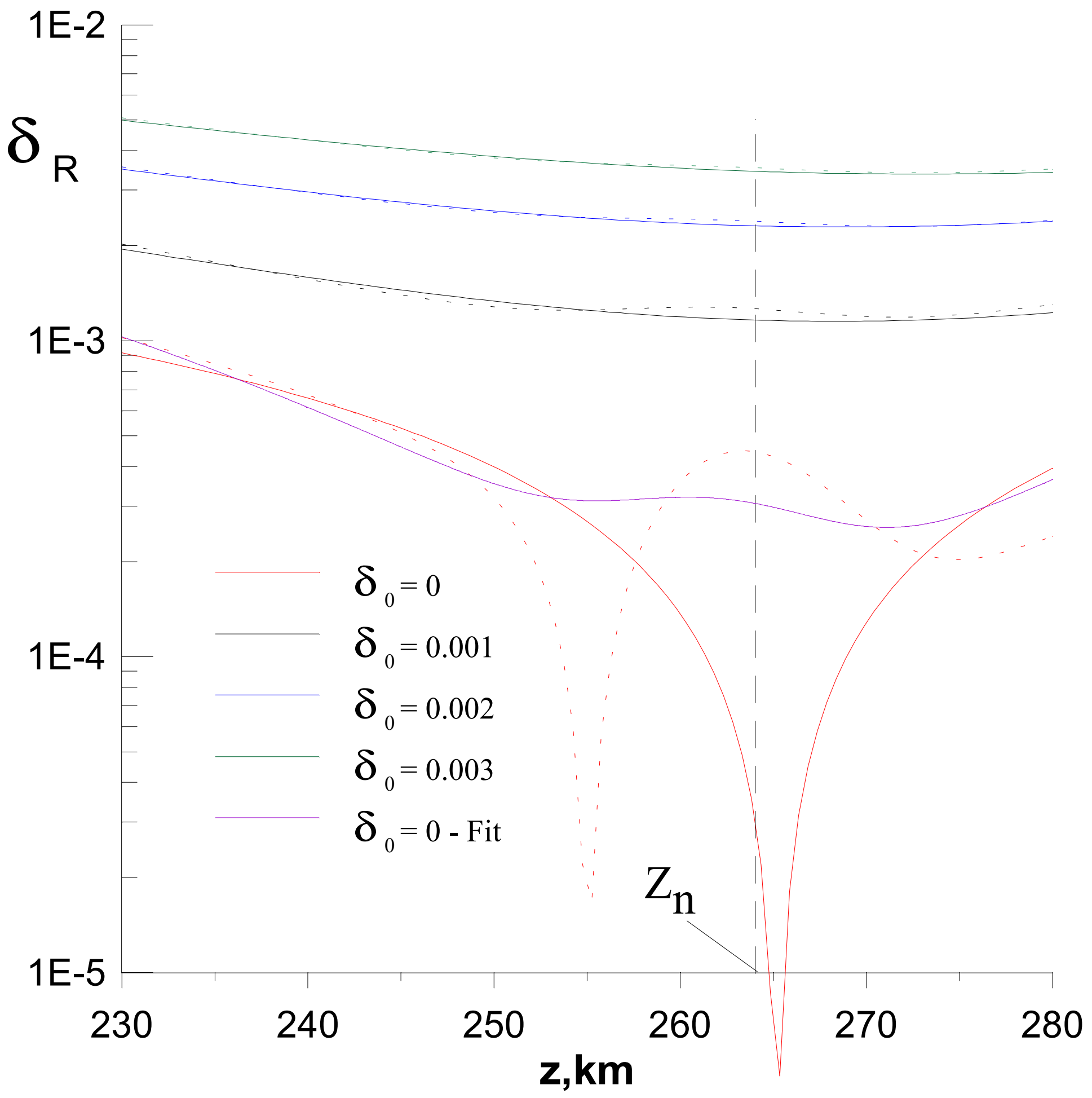

Fig. 12.
Restoration of altitude dependence of $\delta_R$ for the model of parabolic profile with a "hump" at the height of reflection of pump wave. Experimental data are for September, 7, 1999.

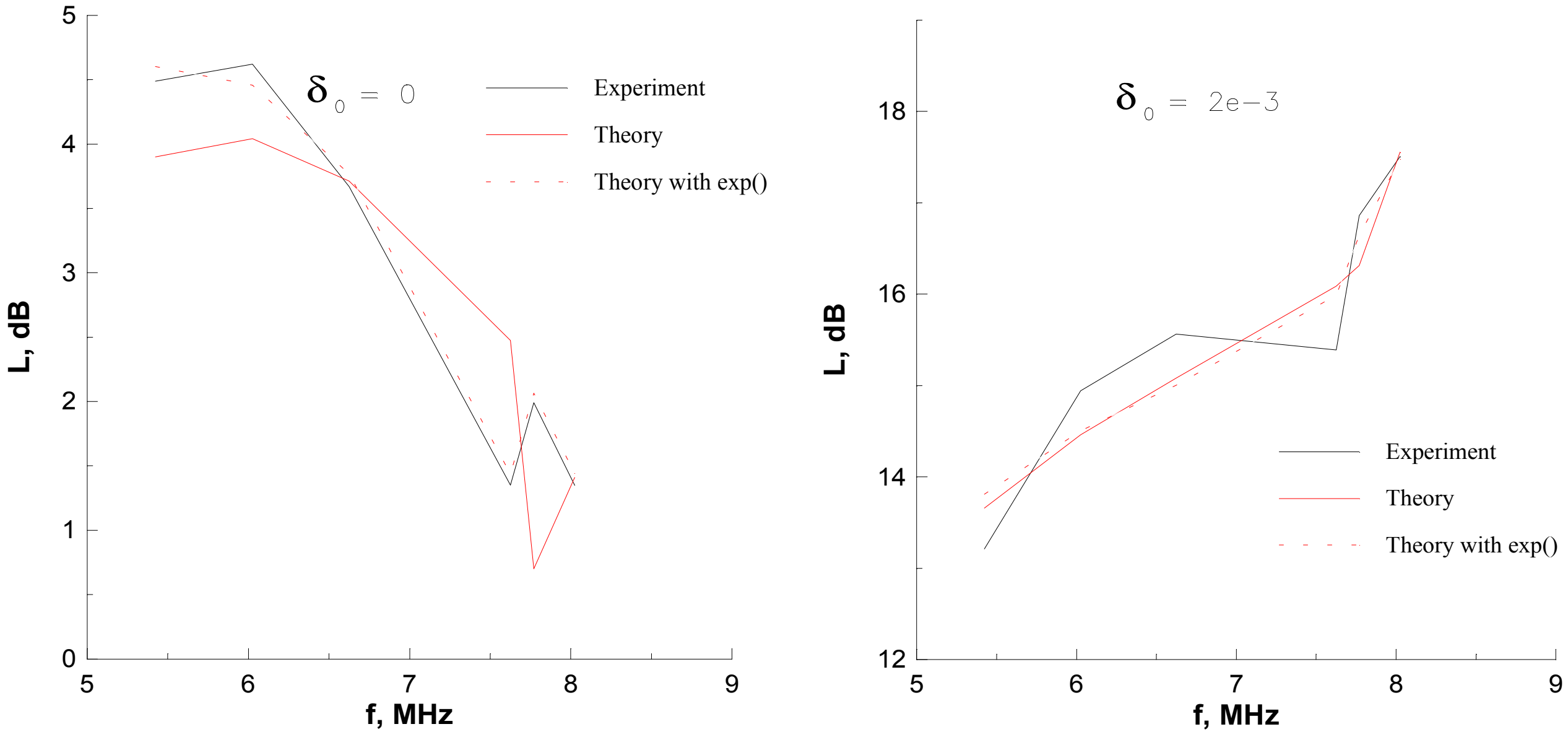

Fig. 13.
Comparison of anomalous attenuation calculated for parabolic model with the "hump" with experimental data for September, 7, 1999.